\documentclass[twocolumn,english,aps,prl,showpacs,superscriptaddress]{revtex4-1}
\usepackage[T1]{fontenc}
\usepackage[latin9]{inputenc}
\usepackage{amsmath}
\usepackage{amssymb}
\usepackage{graphicx}

\makeatletter

\@ifundefined{textcolor}{}
{%
 \definecolor{BLACK}{gray}{0}
 \definecolor{WHITE}{gray}{1}
 \definecolor{RED}{rgb}{1,0,0}
 \definecolor{GREEN}{rgb}{0,1,0}
 \definecolor{BLUE}{rgb}{0,0,1}
 \definecolor{CYAN}{cmyk}{1,0,0,0}
 \definecolor{MAGENTA}{cmyk}{0,1,0,0}
 \definecolor{YELLOW}{cmyk}{0,0,1,0}
}

\usepackage{babel}

\makeatother

\usepackage{babel}
\begin{document}

\title{Coherent coupling of alkali atoms by random collisions}

\author{Or Katz}

\affiliation{Department of Physics of Complex Systems, Weizmann Institute of Science,
Rehovot 76100, Israel}

\affiliation{Rafael Ltd, IL-31021 Haifa, Israel}

\author{Or Peleg}

\affiliation{Rafael Ltd, IL-31021 Haifa, Israel}

\author{Ofer Firstenberg}

\email[Corresponding author:]{ofer.firstenberg@weizmann.ac.il}

\affiliation{Department of Physics of Complex Systems, Weizmann Institute of Science,
Rehovot 76100, Israel}
\begin{abstract}
Random spin-exchange collisions in warm alkali vapor cause rapid decoherence
and act to equilibriate the spin state of the atoms. In contrast,
here we demonstrate experimentally and theoretically a \textit{coherent
coupling} of one alkali specie to another specie, mediated by these
random collisions. We show that, the minor specie (potassium) inherits
the magnetic properties of the dominant specie (rubidium), including
its lifetime ($T_{1}$), coherence time ($T_{2})$, gyromagnetic ratio,
and SERF magnetic-field threshold. We further show that this coupling
can be completely controlled by varying the strength of the magnetic
field. Finally, we explain these phenomena analytically by modes-mixing
of the two species via spin-exchange collisions.
\end{abstract}
\maketitle
Collisions in a warm alkali vapor relax the quantum spin-state of
the atoms. These collisions usually consist of a few spin-destruction
collisions, limiting the lifetime ($T_{1}$), and of rapid spin-exchange
collisions, limiting coherence times ($T_{2}$) \cite{happer 1972}.
Reduced coherence times affect the performance of many vapor physics
applications, such as atomic clocks \cite{atomic clocks}, optical
memories \cite{storage - lukin,storage images}, and high-field magnetometers
\cite{magnetometers}. However, the spin-exchange (SE) relaxation
can be completely eliminated at a regime known as spin-exchange relaxation-free
(SERF) \cite{happer 1977,Nonlinear SERF}, allowing the realization
of vapor-based ultra-sensitive magnetometers \cite{Romalis SERF magnetometer,SERF magnetometer Kitching}.
At other regimes, SE collisions can be utilized to incoherently transfer
polarization between different atomic states \cite{He3 polarization,happer 1998},
as is widely used in the production of hyper-polarized noble gases
for medical imaging \cite{medical imaging}, spin-polarized targets
\cite{neutron targets}, and precision measurements \cite{CPT1,CPT3}.
The incoherent transfer of polarization via SE collisions was also
demonstrated in a hybrid system of two alkali species \cite{hybrid SEOP,hybrid SEOP 2},
improving magnetometery sensitivity by optically pumping one specie
at a given optical-depth and monitoring the other \cite{romalis hybrid OD}.
However, the \emph{coherent} effects of the SE collisional interaction
between different species of alkali atoms have never been investigated.

In this Letter, we demonstrate a coherent coupling of two alkali species,
potassium (K) and rubidium (Rb), induced by random SE collisions.
The strength of the coupling is controlled by the strength of the
magnetic field. We show that at small magnetic fields, the coupling
is strong and the K atoms inherit the magnetic properties of the Rb,
including its gyromagnetic ratio ($g$) and long coherence-time ($T_{2}$).
Furthermore, the lifetime ($T_{1}$) of the K atoms become that of
the Rb, and the SERF magnetic-field threshold is shown to improve
by an order-of-magnitude, corresponding to that of Rb. We analytically
explain these phenomena by SE-induced hybridization of the quantum
state of the two species.

When two alkali species co-exist in a vapor cell, each atom collides
with atoms of its own specie (S) at a rate $R_{{\scriptscriptstyle \textsc{\textnormal{SE}}}}^{\mbox{\textsc{\textnormal{s-s}}}}$
and with atoms of the other specie (S') at a rate $R_{{\scriptscriptstyle \textsc{\textnormal{SE}}}}^{\mbox{\textsc{\textnormal{s-s'}}}}$.
Achieving strong coupling between the species requires that $R_{{\scriptscriptstyle \textsc{\textnormal{SE}}}}^{\mbox{\textsc{\textnormal{s-s'}}}}$
governs the dynamics of S, exceeding both $R_{{\scriptscriptstyle \textsc{\textnormal{SE}}}}^{\mbox{\textsc{\textnormal{s-s}}}}$
and the Larmor precession rate $\omega_{{\scriptscriptstyle \textsc{\textnormal{S}}}}$.
The rate $R_{{\scriptscriptstyle \textsc{\textnormal{SE}}}}^{\mbox{\textsc{\textnormal{s-s'}}}}=n_{\textsc{\textnormal{s'}}}\sigma_{{\scriptscriptstyle \textsc{\textnormal{SE}}}}^{\textsc{\textnormal{s-s'}}}\bar{v}_{T}$
depends on the mean thermal velocity $\bar{v}_{T}$, the alkali-alkali
SE cross-section $\sigma_{{\scriptscriptstyle \textsc{\textnormal{SE}}}}^{{\scriptscriptstyle \textsc{\textnormal{S-S'}}}}$,
and the density $n_{\textsc{\textnormal{s'}}}$. For Rb and K, the
mutual cross-section $\sigma_{{\scriptscriptstyle \textsc{\textnormal{SE}}}}^{{\scriptscriptstyle \textsc{\textnormal{Rb-K}}}}\approx2\cdot10^{-14}\mbox{\ cm\ensuremath{^{-2}}}$
is similar to the self cross-sections $\sigma_{{\scriptscriptstyle \textsc{\textnormal{SE}}}}^{{\scriptscriptstyle \textsc{\textnormal{K-K}}}}=1.5\cdot10^{-14}\mbox{\ cm\ensuremath{^{-2}}}$
and $\sigma_{{\scriptscriptstyle \textsc{\textnormal{SE}}}}^{{\scriptscriptstyle \textsc{\textnormal{Rb-Rb}}}}=1.9\cdot10^{-14}\mbox{\ cm\ensuremath{^{-2}}}$
\cite{explanation}. When the two alkali drops are not mixed, the
Rb vapor density $n_{{\scriptscriptstyle \textsc{\textnormal{Rb}}}}$
is $10$ times higher than the K density $n_{{\scriptscriptstyle \textsc{\textnormal{K}}}}$\cite{vapor pressure}.
That is, the SE rate experienced by the K is predominantly determined
by the collisions with Rb, $R_{{\scriptscriptstyle \textsc{\textnormal{SE}}}}^{{\scriptscriptstyle \textsc{\textnormal{K-Rb}}}}\sim10R_{{\scriptscriptstyle \textsc{\textnormal{SE}}}}^{{\scriptscriptstyle \textsc{\textnormal{K-K}}}}$,
whereas the Rb is only weakly affected by the presence of the K since
$R_{{\scriptscriptstyle \textsc{\textnormal{SE}}}}^{{\scriptscriptstyle \textsc{\textnormal{Rb-Rb}}}}\sim10R_{{\scriptscriptstyle \textsc{\textnormal{SE}}}}^{{\scriptscriptstyle \textsc{\textnormal{Rb-K}}}}$.
Thus, at high SE rates the K dynamics will be significantly affected
by the presence of the Rb.

\begin{figure}[b]
\includegraphics[width=7.3cm]{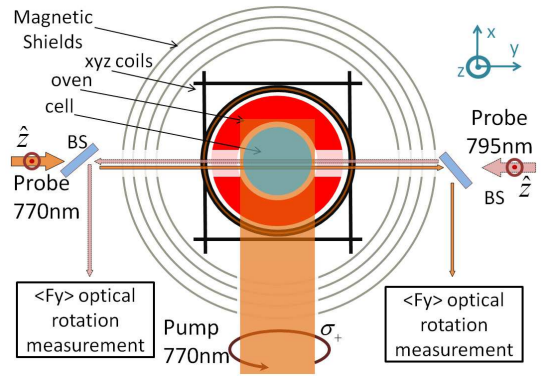} \protect\caption{Schematics of the experimental setup. The K atoms are initially polarized
by a $\sigma_{+}\mbox{-polarized}$ pumping beam. The polarization
rotation of two linearly-polarized beams ($770\,\mbox{nm}$ and $795\mbox{\ nm}$)
is measured to probe the atomic spin precession of both K and Rb atoms,
after a magnetic field $B_{z}$ is applied. BS: non-polarizing beam
splitter. }

\label{fig:Experimental-setup} 
\end{figure}

To probe the magnetic coherence of the two species we use a single
pump and two probe beams as shown in Fig.~\ref{fig:Experimental-setup}.
We use a spherical cell of radius $0.5"$ with natural abundance of
Rb ($\ensuremath{72\%}$ $\mbox{\ensuremath{^{85}}Rb}$ and $\ensuremath{28\%}$
$\mbox{\ensuremath{^{87}}Rb}$) and K, in two separate drops. The
cell is heated to a temperature of $T=95{}^{\circ}\mbox{C}$ , corresponding
to number densities of $n_{{\scriptscriptstyle \textsc{\textnormal{Rb}}}}\sim4\cdot10^{12}\;\mbox{cm\ensuremath{^{-3}}}$
and $n_{{\scriptscriptstyle \textsc{\textnormal{K}}}}\sim4\cdot10^{11}\;\mbox{cm\ensuremath{^{-3}}}$.
We reduce radiation trapping and collision rate with the cell walls
by introducing $200\;\mbox{torr}$ of $\mbox{\ensuremath{N_{2}}}$
buffer gas. The magnetic field is controlled by shielding the cell
with $\mu\mbox{-metal}$ cylinders and three perpendicular Helmholtz
coils. We optically pump the K using a circularly-polarized laser
beam at $770\;\mbox{nm}$ along the $\hat{x}$ direction. The beam
is wider than the cell with intensity of $1\;\mbox{mW/cm\ensuremath{^{2}}}$,
tuned to the $D1$ absorption peak. During the pumping, the K atoms
pump the Rb atoms via SE collisions. We then switch off the pumping
light and apply a magnetic field $B\hat{z}$. Consequently, both Rb
and K spins precess around the magnetic field while decaying due to
the various relaxation mechanisms in the cell. We monitor this precession
by measuring the optical rotation of linearly-polarized probe beams
at $770\mbox{\ nm}$ and $795\;\mbox{nm}$, close to resonance with
the K and Rb, respectively. We ensure that the probes do not induce
relaxation by using moderate intensities ($I_{{\scriptscriptstyle \textsc{\textnormal{Rb}}}}=1\;\mbox{mW/cm\ensuremath{^{2}}}$,
$I_{{\scriptscriptstyle \textsc{\textnormal{K}}}}=10\;\mbox{mW/cm\ensuremath{^{2}}}$)
and detuning from the $D1$ line {[}$\Delta\nu_{{\scriptscriptstyle Rb}}=25\,\mbox{\ensuremath{\left(2\pi\right)}GHz}$,
$\Delta\nu_{{\scriptscriptstyle \textsc{\textnormal{K}}}}=15\,\mbox{\ensuremath{\left(2\pi\right)}GHz}${]}.
As a reference, we measured the magnetic coherence of K in a similar
cell, without Rb.

\begin{figure}[t]
\centering\includegraphics[width=8.6cm,height=4.3cm]{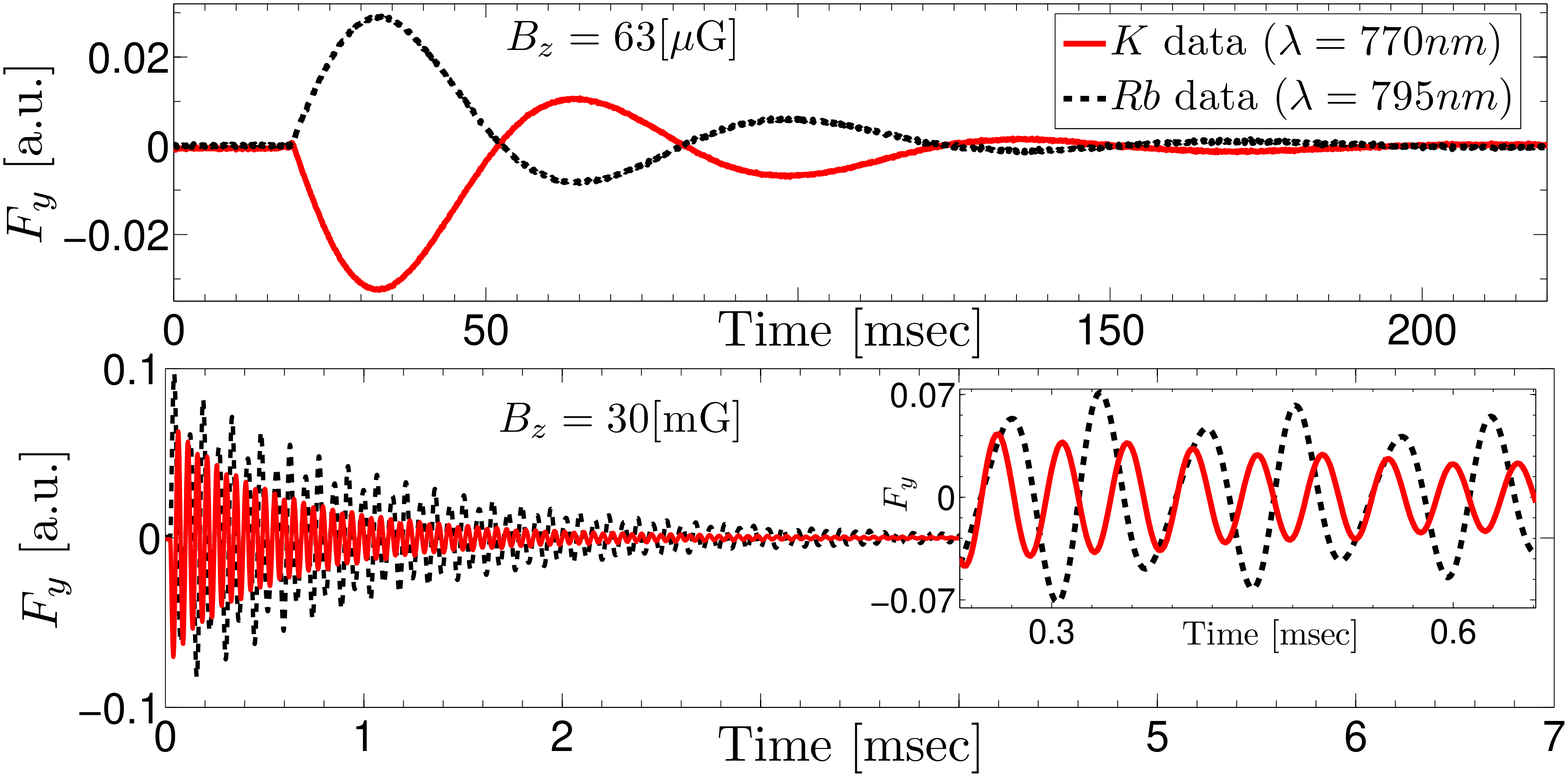} \protect\caption{Typical measurement of $F_{y}$ (the $\hat{y}$ component of the spin)
for K and Rb. At low magnetic fields (top), the two species are coherently
coupled- oscillate and decay at the same rate, while at high magnetic
fields (bottom), they are decoupled- oscillate and decay differently.}

\label{fig:typical oscillation} 
\end{figure}

\begin{figure}[t]
\centering\includegraphics[bb=0bp 0bp 990bp 532bp,clip,width=8.6cm]{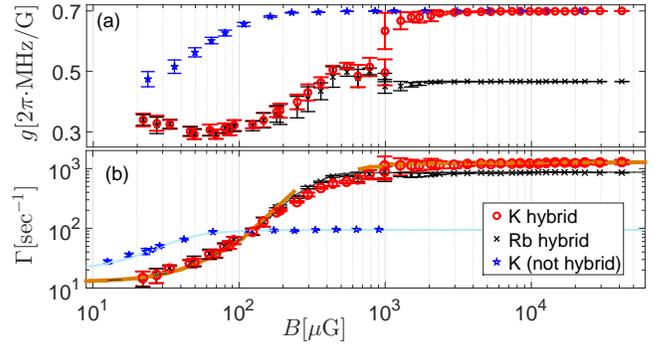}
\protect\caption{(a). Measured gyromagnetic ratios $g=\omega/B_{z}$ of Rb (black)
and K (red) in the hybrid cell. The K inherits the precession rate
of Rb at low magnetic fields. The measured $g$ of K in a reference
cell without Rb is shown for comparison (blue). (b) Measured decoherence
rate $\Gamma=1/T_{2}$ of the two species. $\Gamma_{{\scriptscriptstyle \textsc{\textnormal{K}}}}$
is significantly improved at low magnetic fields. Both species experience
SERF at the same magnetic-field threshold, 10 times higher than the
threshold of K without Rb. The solid lines are the analytically-calculated
values. }

\label{fig:decay and gyro-magnetic} 
\end{figure}

A typical measurement of the optical rotation signals at low and high
magnetic fields is shown in Fig.~\ref{fig:typical oscillation}.
At high magnetic fields, the two species are weakly coupled and oscillate
at different frequencies $\omega_{{\scriptscriptstyle \textsc{\textnormal{K}}}}=1.5\cdot\omega_{{\scriptscriptstyle \textsc{\textnormal{Rb}}}}$,
corresponding to their natural gyromagnetic ratios \cite{happer 1977},
\begin{equation}
g=\frac{\mu_{B}g_{s}}{\hbar}\frac{1}{\left(2I+1\right)},\label{eq:gyro_magnetic_ratio}
\end{equation}
where $I_{{\scriptscriptstyle \textsc{\textnormal{\ensuremath{\mbox{\ensuremath{^{85}}}\mbox{Rb}}}}}}=5/2$,
$I_{{\scriptscriptstyle \textsc{\textnormal{K}}}}=3/2$, and $g_{s}=2$.
However at low magnetic fields, the two species are strongly coupled
and thus precess at the same frequency and have the same decoherence
rate. The frequencies and decoherence rates were determined by fitting
each signal to the simple model $f=\sum_{i=1,2}c_{i}e^{-\Gamma_{i}t}\sin(\omega_{i}t+\varphi_{i})$
assuming that $c_{i},\Gamma_{i},\omega_{i},$and $\varphi_{i}$ are
constants. We note that the $180^{\circ}$ phase between the signals
results from the counter-propagating arrangement of the probing beams.
The beating of the Rb signal (Fig.~\ref{fig:typical oscillation}
bottom) originated from the presence of the $\mbox{\ensuremath{^{87}}Rb}$
isotope, and its effect on the $K$ dynamics is discussed in the supplemental
material \cite{supplemnetary-1}. At low and at high magnetic fields,
each signal has a single dominant precession rate $\omega_{i}$ with
a corresponding decoherence rate $\Gamma_{i}$. Therefore in these
regimes, we identify for each specie the dominant precession rate
$\omega_{{\scriptscriptstyle \textsc{\textnormal{K}}}}$, $\omega_{{\scriptscriptstyle \textsc{\textnormal{Rb}}}}$
and decoherence rate $\Gamma_{{\scriptscriptstyle \textsc{\textnormal{K}}}}$,
$\Gamma_{{\scriptscriptstyle \textsc{\textnormal{Rb}}}}$, shown in
Fig.~\ref{fig:decay and gyro-magnetic}. 


The \emph{coherent} coupling is demonstrated in Fig.~\ref{fig:decay and gyro-magnetic}a
by examining the precession rates of the two species. While the gyromagnetic
ratio of the Rb is practically unaffected by the presence of the K
(corresponding to its normal SERF behavior \cite{happer 1977,Nonlinear SERF}),
the K dynamics is strongly influenced by the Rb. At high magnetic
fields, the K is well decoupled and precesses at its natural rate,
$g=0.7\:\mbox{\ensuremath{\left(2\pi\right)}MHz/G}$. However, at
low magnetic fields, the coupling is strong and the K precesses at
the gyromagnetic ratio of Rb, $g=0.26\,\mbox{\ensuremath{\left(2\pi\right)}MHz/G}$,
and not at its natural low-field precession rate \cite{happer 1977,Nonlinear SERF},
$g=0.46\,\mbox{\ensuremath{\left(2\pi\right)}MHz/G}$.

\begin{figure}[t]
\includegraphics[clip,width=8.6cm,height=4cm]{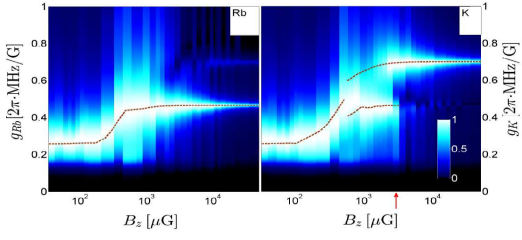}
\protect\caption{Spectral content of the measured signals as a function of the magnetic
field. The vertical axis is scaled to measure the gyromagnetic ratio.
The Rb precesses at a single rate (left) while the K inherits the
rate of the Rb at low fields. The the theoretical $g$ is plotted
in dashed lines.}

\label{fig:fourier_maps} 
\end{figure}

To investigate the transition between the coupled and uncoupled states
of the two species, we show in Fig.~\ref{fig:fourier_maps} the spectral
dependence of the signals on the applied magnetic field. We plot the
FFT of the measured signals at different magnetic fields using a normalized
scale ($\omega\rightarrow\omega/B_{z}$) corresponding to the gyromagnetic
ratio $g$ (for visualization purposes, we replace with zero the FFT
around $g=0$). At high and low magnetic fields, the physics is exactly
the same as in Fig.~\ref{fig:decay and gyro-magnetic}a: the two
species share the same gyromagnetic ratio (that of the Rb) at low
fields, while reaching their normal different values at high magnetic
fields. However at intermediate fields, the K precesses at a combination
of two frequencies, both at its natural precession rate and at the
rate of $\mbox{\ensuremath{^{85}}Rb}$ (most clearly visible at $B_{z}\mbox{=3 mG}$,
red arrow in Fig.~\ref{fig:fourier_maps}). The nature of this transition
can not be explained by a mere shift in the slowing-down-factor since
the K dynamics consists of two distinct frequencies. Thus, the minor
specie (K) is coherently coupled to the dominant specie (Rb) at low
magnetic fields by SE collisions, and the coupling strength is controlled
by the magnitude of the magnetic field. 

This coherent coupling alters significantly other important properties
of the K. While the relaxation rate $\Gamma_{Rb}\left(B_{z}\right)$
is unaffected by collisions with K (SE relaxation is eliminated at
$\omega_{{\scriptscriptstyle \textsc{\textnormal{Rb}}}}\lesssim R_{{\scriptscriptstyle \textsc{\textnormal{SE}}}}^{{\scriptscriptstyle \textsc{\textnormal{Rb-Rb}}}}$
\cite{happer 1977}, as if the K is absent), the relaxation rate $\Gamma_{K}\left(B_{z}\right)$
is inherited from the Rb, as shown in Fig.~\ref{fig:decay and gyro-magnetic}b.
Consequently, the relaxation rate $\Gamma_{K}\left(B_{z}\right)$
decreases dramatically at low magnetic fields, in comparison to a
cell without Rb. The elimination of relaxation of the K atoms appears
at $\omega_{{\scriptscriptstyle \textsc{\textnormal{K}}}}\lesssim R_{{\scriptscriptstyle \textsc{\textnormal{SE}}}}^{{\scriptscriptstyle \textsc{\textnormal{Rb-K}}}}$,
\textit{i.e.} at magnetic fields 10-times higher (less stringent)
than those required without the Rb, $\omega_{{\scriptscriptstyle \textsc{\textnormal{K}}}}\lesssim R_{{\scriptscriptstyle \textsc{\textnormal{SE}}}}^{{\scriptscriptstyle \textsc{\textnormal{K-K}}}}$.
Thus, the dominant specie (Rb) drives the minor specie (K) into SERF
at the magnetic-field threshold of the dominant specie. At high magnetic
fields, the relaxation of the minor specie is increased due to mutual
SE collisions. Furthermore, we measure an identical lifetime of both
species ($T_{1}=86\pm1\;\mbox{msec}$). Via collisions, the K inherits
the lifetime of the Rb incoherently, as theoretically explained for
different isotopes in \cite{happer 1998}. Since the spin destruction
is dominated by electron-randomizing collisions (\textit{e.g.,} anisotropic
collisions with $\mbox{N}_{2}$), the K benefits from the larger nuclear
spin reservoir of the Rb. Therefore, through collisional induced coherent
coupling, the minor specie (K) inherits the magnetic properties of
the dominant specie (Rb), including its precession rate, decoherence
rate, lifetime, and SERF magnetic field threshold.

To understand the nature of the coherent coupling, we solve the coupled
Liouville equations describing the dynamics of the system. Due to
frequent SE collisions, the coherence between the two hyperfine levels
decays rapidly. Therefore, the ground-state density matrix of each
specie $s$ can be simplified by decomposing $\rho^{{\scriptscriptstyle \textsc{\textnormal{S}}}}$
into a two block diagonal form $\rho^{{\scriptscriptstyle \textsc{\textnormal{S}}}}=\rho_{a}^{{\scriptscriptstyle \textsc{\textnormal{S}}}}+\rho_{b}^{{\scriptscriptstyle \textsc{\textnormal{S}}}}$,
where $a=I_{{\scriptscriptstyle \textsc{\textnormal{S}}}}+1/2$ denotes
the higher hyperfine level and $b=I_{{\scriptscriptstyle \textsc{\textnormal{S}}}}-1/2$
the lower one \cite{SERF high polarization-1}. The Larmor coherence
can be described by the transverse spin vector $\langle F_{+}^{{\scriptscriptstyle \textsc{\textnormal{S}}}}\rangle\equiv-F_{x}^{{\scriptscriptstyle \textsc{\textnormal{S}}}}+iF_{y}^{{\scriptscriptstyle \textsc{\textnormal{S}}}}=F_{a+}^{{\scriptscriptstyle \textsc{\textnormal{S}}}}+F_{b+}^{{\scriptscriptstyle \textsc{\textnormal{S}}}}$,
where we omit the brackets $\langle\rangle$ for brevity. In the absence
of the mutual SE interaction, the dynamics of $F_{+}^{{\scriptscriptstyle \textsc{\textnormal{S}}}}$
for each specie is given by \cite{supplemnetary-1}

\begin{align}
\frac{d}{dt}\left(\begin{array}{c}
F_{a+}^{{\scriptscriptstyle \textsc{\textnormal{S}}}}\\
F_{b+}^{{\scriptscriptstyle \textsc{\textnormal{S}}}}
\end{array}\right) & =\left(-ig_{{\scriptscriptstyle \textsc{\textnormal{S}}}}B_{z}\left(\begin{array}{cc}
1 & 0\\
0 & -1
\end{array}\right)-R_{{\scriptscriptstyle \textsc{\textnormal{SE}}}}^{\mbox{\textsc{\textnormal{s-s}}}}\left(\begin{array}{cc}
x_{a}^{{\scriptscriptstyle \textsc{\textnormal{S}}}} & y^{{\scriptscriptstyle \textsc{\textnormal{S}}}}\\
y^{{\scriptscriptstyle \textsc{\textnormal{S}}}} & x_{b}^{{\scriptscriptstyle \textsc{\textnormal{S}}}}
\end{array}\right)\right.\nonumber \\
- & \left.R_{{\scriptscriptstyle \textsc{\textnormal{SD}}}}^{\mbox{\textsc{\textnormal{s-s}}}}\left(\begin{array}{cc}
w_{a}^{{\scriptscriptstyle \textsc{\textnormal{S}}}} & z^{{\scriptscriptstyle \textsc{\textnormal{S}}}}\\
z^{{\scriptscriptstyle \textsc{\textnormal{S}}}} & w_{b}^{{\scriptscriptstyle \textsc{\textnormal{S}}}}
\end{array}\right)\right)\left(\begin{array}{c}
F_{a+}^{{\scriptscriptstyle \textsc{\textnormal{S}}}}\\
F_{b+}^{{\scriptscriptstyle \textsc{\textnormal{S}}}}
\end{array}\right),\label{eq:LINERIZED SELF SE}
\end{align}
where $g_{{\scriptscriptstyle \textsc{\textnormal{S}}}}$ is the gyromagnetic
ratio of each specie {[}Eq.~\eqref{eq:gyro_magnetic_ratio}{]}, $R_{{\scriptscriptstyle \textsc{\textnormal{SD}}}}^{\mbox{\textsc{\textnormal{s-s}}}}$
is the spin-destruction rate, and the constant coefficients $w_{a}^{{\scriptscriptstyle \textsc{\textnormal{S}}}},w_{b}^{{\scriptscriptstyle \textsc{\textnormal{S}}}},x_{a}^{{\scriptscriptstyle \textsc{\textnormal{S}}}},x_{b}^{{\scriptscriptstyle \textsc{\textnormal{S}}}},y^{{\scriptscriptstyle \textsc{\textnormal{S}}}},z^{{\scriptscriptstyle \textsc{\textnormal{S}}}}$
are given in \cite{supplemnetary-1}, depending on the nuclear spin
$I_{{\scriptscriptstyle \textsc{\textnormal{S}}}}$ only. The SE coupling
between the two species supplements Eq.~\eqref{eq:LINERIZED SELF SE}
by inducing the coupling interaction 
\begin{equation}
\frac{d}{dt}\left(\mathbf{\mathbf{\tilde{F}_{+}}}\right)_{coupling}=-\overleftrightarrow{M}\cdot\mathbf{\tilde{F}_{+}},\label{eq:coupling term}
\end{equation}
where $\mathbf{\tilde{F}_{+}}$ denotes the 4-component vector $\mathbf{\tilde{F}_{+}}\equiv\left(F_{a+}^{{\scriptscriptstyle \textsc{\textnormal{Rb}}}},F_{b+}^{{\scriptscriptstyle \textsc{\textnormal{Rb}}}},F_{a+}^{{\scriptscriptstyle \textsc{\textnormal{K}}}},F_{b+}^{{\scriptscriptstyle \textsc{\textnormal{K}}}}\right)$
and $\overleftrightarrow{M}$ is the $4\times4$ linearized SE coupling
matrix, given by 
\begin{equation}
\overleftrightarrow{M}=diag\left(\begin{array}{c}
R_{{\scriptscriptstyle \textsc{\textnormal{SE}}}}^{\mbox{\ensuremath{{\scriptscriptstyle \textsc{\textnormal{Rb-K}}}}}}\\
R_{{\scriptscriptstyle \textsc{\textnormal{SE}}}}^{\mbox{\ensuremath{{\scriptscriptstyle \textsc{\textnormal{Rb-K}}}}}}\\
R_{{\scriptscriptstyle \textsc{\textnormal{SE}}}}^{\mbox{\ensuremath{{\scriptscriptstyle \textsc{\textnormal{K-Rb}}}}}}\\
R_{{\scriptscriptstyle \textsc{\textnormal{SE}}}}^{\mbox{\ensuremath{{\scriptscriptstyle \textsc{\textnormal{K-Rb}}}}}}
\end{array}\right)\cdot\left(\begin{array}{cccc}
w_{a}^{{\scriptscriptstyle \textsc{\textnormal{Rb}}}} & z^{{\scriptscriptstyle \textsc{\textnormal{Rb}}}} & G_{11}^{{\scriptscriptstyle \textsc{\textnormal{Rb}}}} & G_{12}^{{\scriptscriptstyle \textsc{\textnormal{Rb}}}}\\
z^{{\scriptscriptstyle \textsc{\textnormal{Rb}}}} & w_{b}^{{\scriptscriptstyle \textsc{\textnormal{Rb}}}} & G_{21}^{{\scriptscriptstyle \textsc{\textnormal{Rb}}}} & G_{22}^{{\scriptscriptstyle \textsc{\textnormal{Rb}}}}\\
G_{11}^{{\scriptscriptstyle \textsc{\textnormal{K}}}} & G_{21}^{{\scriptscriptstyle \textsc{\textnormal{K}}}} & w_{a}^{{\scriptscriptstyle \textsc{\textnormal{K}}}} & z^{{\scriptscriptstyle \textsc{\textnormal{K}}}}\\
G_{12}^{{\scriptscriptstyle \textsc{\textnormal{K}}}} & G_{22}^{{\scriptscriptstyle \textsc{\textnormal{K}}}} & z^{{\scriptscriptstyle \textsc{\textnormal{K}}}} & w_{b}^{{\scriptscriptstyle \textsc{\textnormal{K}}}}
\end{array}\right).\label{eq:coupling matrix-1}
\end{equation}
Here, the constants $G_{11}^{{\scriptscriptstyle \textsc{\textnormal{S}}}},G_{12}^{{\scriptscriptstyle \textsc{\textnormal{S}}}},G_{21}^{{\scriptscriptstyle \textsc{\textnormal{S}}}},G_{22}^{{\scriptscriptstyle \textsc{\textnormal{S}}}}$
depend on $I_{{\scriptscriptstyle \textsc{\textnormal{Rb}}}}$ and
$I_{{\scriptscriptstyle \textsc{\textnormal{K}}}}$ only \cite{supplemnetary-1}.
Qualitatively, this coupling interaction has two major effects on
the dynamics of the vapor. First, it increases the effective spin-destruction
rate of the two species, by $R_{{\scriptscriptstyle \textsc{\textnormal{SE}}}}^{\mbox{\ensuremath{{\scriptscriptstyle \textsc{\textnormal{Rb-K}}}}}}$
for the Rb and $R_{{\scriptscriptstyle \textsc{\textnormal{SE}}}}^{\mbox{\ensuremath{{\scriptscriptstyle \textsc{\textnormal{K-Rb}}}}}}$
for the K, as can be seen by the block-diagonal elements in Eq.~\eqref{eq:coupling matrix-1}.
Second, since $R_{{\scriptscriptstyle \textsc{\textnormal{SE}}}}^{\mbox{\ensuremath{{\scriptscriptstyle \textsc{\textnormal{K-Rb}}}}}}\gg R_{{\scriptscriptstyle \textsc{\textnormal{SE}}}}^{\mbox{\ensuremath{{\scriptscriptstyle \textsc{\textnormal{K-K}}}}}},R_{{\scriptscriptstyle \textsc{\textnormal{SD}}}}^{\mbox{\ensuremath{{\scriptscriptstyle \textsc{\textnormal{K-K}}}}}}$,
the coupling through the coefficients $G_{ij}^{{\scriptscriptstyle \textsc{\textnormal{K}}}}$
becomes the dominant collisional interaction for the K, while the
coupling experienced by the Rb is weak, since $R_{{\scriptscriptstyle \textsc{\textnormal{SE}}}}^{\mbox{\ensuremath{{\scriptscriptstyle \textsc{\textnormal{Rb-K}}}}}}\ll R_{{\scriptscriptstyle \textsc{\textnormal{SE}}}}^{\mbox{\ensuremath{{\scriptscriptstyle \textsc{\textnormal{Rb-Rb}}}}}}$.
At low magnetic fields, $R_{{\scriptscriptstyle \textsc{\textnormal{SE}}}}^{\mbox{\ensuremath{{\scriptscriptstyle \textsc{\textnormal{K-Rb}}}}}}$
is the dominant rate for the K spins, which therefore respond to the
precession of the Rb spins as forced oscillations and inherit the
Rb magnetic properties. At the same time, the Rb dynamics remain unaffected
by the K.

\begin{figure}[t]
\centering\includegraphics[clip,width=8.6cm]{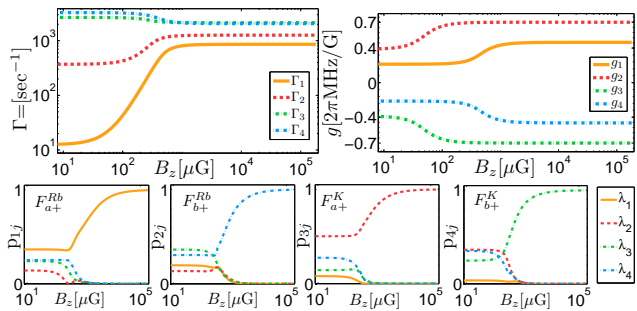}

\protect\caption{Theoretical results. Top: Decoherence rates $-\mbox{Re}\left(\lambda\right)$
(left) and gyromagnetic ratios $\mbox{Im}(\lambda/B_{z})$ (right)
of the four eigenmodes. The decoherence rate $\Gamma_{1}$ vanishes
at low magnetic fields due to SERF. Bottom: The coupling coefficients
$p_{ij}$ of the four spin components. At high magnetic fields, the
spins of K and $\mbox{\ensuremath{^{85}}}\mbox{Rb}$ consist of a
single mode; at low fields, the spins consist of all modes, and thus
share their coherence.}

\label{fig:coefficients} 
\end{figure}

To quantitatively describe this coupling, we solve the two coupled
Eqs.~\eqref{eq:LINERIZED SELF SE}-\eqref{eq:coupling term} by diagonalizing
the $4\times4$ matrix and calculating the eigenvalues $\lambda_{i}$
of the four eigenmodes $|\lambda_{i}\rangle$. Each spin component
of $\mathbf{\tilde{F}}_{+}^{i}$ can be spanned by the eigenmodes,
\begin{equation}
\mathbf{\tilde{F}}_{+}^{i}\left(t\right)=\mathbf{\tilde{F}}_{+}^{i}\left(0\right)\sum_{j=1}^{4}p_{ij}e^{\lambda_{j}t},\label{eq:linear combination of modes}
\end{equation}
where the coefficients $p_{ij}$ satisfy the normalization condition
$\sum_{j}p_{ij}=1$. The decoherence rates and gyromagnetic ratios
of the eigenmodes, corresponding to $\Gamma_{i}=-\mbox{Re}(\lambda_{i})$
and $g_{i}=\mbox{Im}(\lambda_{i}/B_{z})$ respectively, are shown
in Fig.~\ref{fig:coefficients} (top). At high magnetic fields, the
gyromagnetic ratios $g_{1}=-g_{4}=0.47\,\mbox{\ensuremath{\left(2\pi\right)}MHz/G}$
correspond to the precession rate of $\mbox{\ensuremath{^{85}}Rb}$,
while $g_{2}=-g_{3}=0.7\,\mbox{\ensuremath{\left(2\pi\right)}MHz/G}$
correspond to the rate of K {[}Eq.~\eqref{eq:gyro_magnetic_ratio}{]}.
The sign of $g$ corresponds to the co-rotation of the upper hyperfine
level ($a$) for a positive value and counter-rotation of the lower
level ($b$) for a negative value. $\Gamma_{1}$ remains the lowest
decay rate of the system at all magnetic fields. As the magnetic field
is lowered, $\Gamma_{1}\left(B\right)$ decreases (experiencing SERF),
and consequently $|\lambda_{1}\rangle$ becomes the dominant eigenmode. 

To infer the dynamics of the atoms, we plot the coefficients $p_{ij}$
of each specie in Fig.~\ref{fig:coefficients} (bottom). At high
magnetic fields, the eigenmodes are well resolved, corresponding to
$|\lambda_{1}\rangle$ for $F_{a+}^{{\scriptscriptstyle \textsc{\textnormal{Rb}}}}$,
$|\lambda_{2}\rangle$ for $F_{a+}^{{\scriptscriptstyle \textsc{\textnormal{K}}}}$,
$|\lambda_{3}\rangle$ for $F_{b+}^{{\scriptscriptstyle \textsc{\textnormal{K}}}}$,
and $|\lambda_{4}\rangle$ for $F_{b+}^{{\scriptscriptstyle \textsc{\textnormal{Rb}}}}$.
Consequently, the different hyperfine levels remain resolved, such
that each magnetic coherence rotates at its natural rate $g_{i}$.
As the magnetic field is lowered, $\mathbf{\tilde{F}}_{+}^{i}$ becomes
a mixture of the four different modes, inducing mixing between the
different hyperfine levels of both species. Since at low magnetic
fields $\Gamma_{1}\ll\Gamma_{2},\Gamma_{3},\Gamma_{4}$, eigenmode
$|\lambda_{1}\rangle$ dominates; The K and Rb spins precess and decay
at the same rate and experience SERF at the Rb's transition rate.
This common effectively-single eigenmode shows that, at low magnetic
fields, the two species hybridize by the SE interaction. At intermediate
magnetic fields, $\Gamma_{1}$ increases and more eigenmodes become
important, resulting in multiple decoherence and precession rates
for each specie, especially for the minor one (K). At this transition
regime, the high decoherence rates $\Gamma_{i}\gtrsim\omega_{i}-\omega_{j}$
broaden the spectral profile of the signals, as observed in Fig.~\ref{fig:fourier_maps}.

To compare with the experiment, the theoretical decoherence rate and
gyromagnetic ratio are plotted in Figs.~\ref{fig:decay and gyro-magnetic}
and \ref{fig:fourier_maps}. These parameters were evaluated from
Eq.~\eqref{eq:linear combination of modes}, while taking into account
the presence of $\mbox{\ensuremath{^{87}}Rb}$ in our experiment \cite{supplemnetary-1},
without any free parameters. $\Gamma_{{\scriptscriptstyle \textsc{\textnormal{Rb}}}}$
is almost unaffected by the presence of K and has a single decoherence
rate and gyromagnetic ratio at any magnetic field. The K is governed
by a single mode at low and high magnetic fields and by multiple modes
at intermediate fields. Therefore at this region, we do not associate
a single $\Gamma_{{\scriptscriptstyle \textsc{K}}}$ and show the
two dominant frequencies in Fig.~\ref{fig:fourier_maps}. Our theoretical
model, appears to fully describe the physics of the interaction and
the hybridization of the species.

The SE-hybridization of different alkali species might be utilized
in magnetometery and quantum-information applications. In magnetometery,
one can benefit from the SERF properties of the dominant specie while
monitoring the state of the minor specie, especially when applied
in hybrid magnetometers schemes \cite{romalis hybrid OD} using a
single laser. For quantum-information applications, the hybridization
could introduce a new mechanism to interface vapor-based quantum memories.

In conclusion, we have shown that random SE collisions coherently
couple two different alkali species. The coupling strength is controlled
by the magnitude of the magnetic field. At low magnetic fields, a
minor specie is strongly coupled to the dominant specie and as a result
inherits the magnetic properties of the dominant specie, such as the
coherence-time and gyromagnetic ratio. At intermediate magnetic fields,
we measure a unique response of the minor specie, consisting of two
precession rates. At high magnetic fields, the two species become
decoupled. We explain these phenomena by showing analytically that,
by lowering the magnetic field, the minor specie starts sharing the
main eigenmode of the dominant specie. This effect may be utilized
in different magnetometery and quantum-information schemes.
\begin{acknowledgments}
We thank Mark Dikopoltsev and Elad Greenfeld for their valuable help
in preparing the cell. This work was supported by the Israel Science
Foundation and the Zumbi Stiftung.\end{acknowledgments}

\end{document}